\documentclass[epj]{svjour}

\usepackage{amsmath}
\usepackage{graphicx}
\usepackage{dcolumn}
\usepackage{bm}
\usepackage{amssymb}
\usepackage{latexsym}
\usepackage{psfrag}
\begin{document}
\title{Holographic dilatonic dark energy model}
\author{Alberto Rozas-Fern\'andez\inst{1}}
\institute{Consejo Superior de
Investigaciones Cient{\'i}ficas, Instituto de F{\'i}sica Fundamental, Serrano 121, 28006 Madrid, Spain.}

\date{Received: date / Revised version: date}

\abstract{
We present a dilatonic description of the holographic dark energy
by connecting the holographic dark energy density with the dilaton
scalar field energy density in a flat Friedmann-Robertson-Walker
universe. We show that this model can describe the observed
accelerated expansion of our universe with the choice $c\geq1$ and
reconstruct the kinetic term as well as the dynamics of the
dilaton scalar field.
\PACS{
      {98.80.-k}{Cosmology},   \and
      {95.36.+x}{Dark energy}
     } % end of PACS codes
} %end of abstract
\maketitle
\section{Introduction}
\label{intro}
The fact that the universe is undergoing an epoch of accelerated
expansion is well supported by recent cosmological observations
from Type Ia supernovae (SN Ia) [1], Cosmic Microwave Background
(CMB) anisotropies measured with the WMAP satellite [2], Large
Scale Structure [3], weak lensing [4] and the integrated
Sach-Wolfe effect [5]. Within the framework of the standard
Friedmann-Robertson-Walker (FRW) cosmology, this present
acceleration requires the existence of a negative pressure fluid,
dubbed dark energy, whose pressure $p_{\Lambda}$ and density
$\rho_{\Lambda}$ satisfy
$\omega_{\Lambda}=p_{\Lambda}/\rho_{\Lambda}<-1/3$. In spite of
this mounting observational evidence, the nature and origin of
dark energy remains unknown and has become a fundamental problem
in theoretical physics and observational cosmology. The
cosmological constant (or vacuum energy) is the most obvious
candidate to address this issue as it complies well with the
cosmological tests at our disposal. However, the well known
problem of the cosmological constant and the coincidence problem
[6] are enough reasons to look for alternatives. Interesting
proposals are the quantum cosmic model [7] and $f(R)$ theories
(see [8] for recent reviews and references therein). Likewise, we
have a plethora of dynamical dark energy models such as
quintessence [9], tachyon [10], phantom[11], quintom [12], etc.
Nevertheless, these scalar field dark energy models are only seen
as an effective description of the underlying theory of dark
energy.

On the other hand, based on the validity of effective local
quantum field theory in a box of size $L$, Cohen et al
\cite{Cohen:1998zx} suggested a relationship between the
ultraviolet (UV) and the infrared (IR) cutoffs  due to the limit
set by the formation of a black hole. The $UV-IR$ relationship
gives an upper bound on the zero point energy density

\begin{equation}\label{eq1}
\rho_{\Lambda}\leq L^{-2}M_{p}^{2},
\end{equation}
where $L$ acts as an IR cutoff and $M_p$ is the reduced Planck
mass in natural units. This means that the maximum entropy in a
box of volume $L^{3}$ is
\begin{equation}\label{eq2}
S_{max}\approx S^{3/4}_{BH},
\end{equation}
being $S_{BH}$ the entropy of a black hole of radius $L$. The
largest $L$ is chosen by saturating the bound in Eq.(\ref{eq1}) so
that we obtain the holographic dark energy density
\begin{equation}
\rho_{\Lambda}= 3c^{2}M^{2}_{p}L^{-2},
\end{equation} where c is a free dimensionless $O(1)$ parameter
and the numeric coefficient is chosen for convenience.
Interestingly, this $\rho_{\Lambda}$ is comparable to the observed
dark energy density $\sim10^{-10}eV^{4}$ if we take $L$ as the
Hubble scale $H^{-1}$ being the Hubble parameter at the present
epoch $H=H_{0}\sim10^{-33}eV$.

 However, Hsu
\cite{Hsu:2004ri} pointed out that this does not lead to an
accelerated universe. This led Li \cite{Li:2004rb} to propose that
the IR cut-off $L$ should be taken as the size of the future event
horizon of the universe
\begin{equation}
R_{\rm eh}(a)\equiv a\int\limits_t^\infty{dt'\over
a(t')}=a\int\limits_a^\infty{da'\over Ha'^2}~,\label{eh}
\end{equation}
where $a$ is the scale factor of the universe.

This allows to construct a satisfactory holographic dark energy
(HDE) model which presents a dynamical view of the dark energy that may provide natural solutions to both dark energy problems as
showed in \cite{Li:2004rb}.
The HDE model has been
tested by various observational data including SNIa \cite{r38},
SNIa+BAO+CMB \cite{holodata,Li:2009zs}, X-ray gas mass fraction of galaxy
clusters \cite{r40}, differential ages of passively evolving galaxies
\cite{r41}, Sandage-Leob test \cite{r42}, and so on \cite{Shen:2004ck}. These analyses
show that the HDE model is consistent
with the observational data, being even mildly favoured over the LCDM \cite{Ma:2007pd}.

 As a matter of fact, a time varying dark energy
gives a better fit than a cosmological constant according to some
analysis of astronomical data coming from type Ia supernovae
\cite{dynamical DE data}. However, it must be stressed that almost
all dynamical dark energy models are settled at the
phenomenological level and the HDE model is no exception in this
respect. Its advantage, when compared to other dynamical dark
energy models, is that the HDE model originates from a fundamental
principle in quantum gravity \cite{holoprin}, and therefore possesses some
features of an underlying theory of dark energy. It is then fair to claim that the simplicity and reasonable nature of HDE provide a more reliable
framework for investigating the problem of DE compared with other models proposed in
the literature. For instance, the coincidence problem is substantially
alleviated in some models of HDE based on the assumption that dark matter and
HDE interact, with a decay of HDE into dark matter.

On the other hand, as is well known, the scalar field models are an effective
description of an underlying theory of dark energy. They are popular
not only because of their mathematical simplicity
and phenomenological richness, but also because they
naturally arise in particle physics including supersymmetric field
theories and string/M theory \footnote{to see, for instance, how quintessence and tachyon models arise quite naturally out of the framework of string
theory, consult \cite{Wetterich:1987fm}
 and \cite{Sen:2002in,Sen:2002nu}, respectively}. However, these fundamental theories
do not predict their potential $V(\phi)$ or kinetic term uniquely. We are interested in the following: if we assume the holographic
dark energy energy scenario as the underlying theory of dark
energy, how a scalar field model can be used to effectively
describe it. Therefore, it is meaningful to reconstruct the $V(\phi)$ or kinetic term kinetic term of a dark
energy model possessing some significant features of the quantum
gravity theory, such as the  HDE model. In order to do that, the procedure is to establish a correspondence between the scalar field and the holographic dark energy by identifying their respective energy densities and then reconstruct the potential (if the scalar field is quintessence or the tachyon, for instance) or the kinetic term (k-essence or the dilaton belong to this class) and the dynamics of the field. In this paper, within the different candidates to play the role of the dark energy, we have chosen the dilaton (when it behaves as a scalar field), as this has emerged
as a possible source of dark energy \cite{Piazza:2004df}. Some work has been done in this
direction. Holographic quintessence and holographic quintom models
have been discussed in \cite{holoquint} and \cite{holoquintom},
respectively, the holographic tachyon model in \cite{holotachyon}
and the holographic kinetic k-essence model in
\cite{holokessence}. Other relevant works can be found in
\cite{otherworksholography}. As stated above, the aim of our work is to construct a
holographic dilatonic model of dark energy, relating the dilaton
scalar-field with the HDE.
The rest of the paper can be outlined as follows. In Sec. 2 we
build the holographic dilatonic model and plot the
kinetic term and the evolution of the dilaton field. The conclusions are drawn in Sec.
3.
\section{Holographic dilatonic model of dark energy}
\label{sec:1}

We consider as a starting point the four-dimensional effective low-energy string action which is generally given by
\begin{multline} \label{staction}
 \mathcal{S}=\int d^{4}x\sqrt{-\bar{g}}\{B_{g}(\phi)\tilde{R}+B_{\phi}^{(0)}(\phi))(\tilde{\nabla}\phi)^{2}\\
-\alpha'[c_{1}^{(1)}B_{\phi}^{(1)}(\phi)(\tilde{\nabla}\phi)^{4}+...]+O(\alpha'^{2})\}
\end{multline} where $\phi$ is the dilaton field that controls the strength
of the string coupling $g_s^2$ via the relation
$g_s^2=e^\phi$.
Here we set $\kappa^2=8\pi G=1$. The low-energy effective string theory generates
higher-order derivative terms coming from
$\alpha'$ and loop corrections
(here $\alpha'$ is related to the string length scale $\lambda_{s}$
via the relation $\alpha'=\lambda_{s}/2\pi$).

In the weak coupling regime ($e^{\phi} \ll 1$) the coupling
functions have the dependence
$B_g \simeq B_\phi^{(0)} \simeq B_\phi^{(1)} \simeq e^{-\phi}$.

We shall work in the context of the so-called runaway dilaton
scenario \cite{Gasperini:2001pc} in which the coupling functions
in Eq.~(\ref{staction}) are given by
\begin{eqnarray}
& &B_g(\phi)=C_g+D_ge^{-\phi}+
{\cal O}(e^{-2\phi})\,, \\
& &B_\phi^{(0)}(\phi)=C_\phi^{(0)}+D_\phi^{(0)}
e^{-\phi}+
{\cal O}(e^{-2\phi})\,.
\end{eqnarray}
In this case $B_g(\phi)$ and $B_\phi^{(0)}(\phi)$
approach constant values as $\phi \to \infty$. Hence the dilaton gradually decouples from gravity
as the field evolves toward the region $\phi \gg 1$
from the weakly coupled regime and we assume that the dilaton is effectively decoupled
from gravity in the limit $\phi \to \infty$ and therefore behaves as a scalar field.

Once we assume that the dilaton behaves as a scalar field, we consider the following general 4-dimensional action

\begin{equation}\label{actiondilaton}
S= S_{grav}+S_{\phi}=\int d^{4}x
\sqrt{-g}\left[\frac{M_{P}^{2}}{2}R+ p_D(X,\phi)\right],
\end{equation}
where $R$ is the Ricci scalar and the effective Lagrangian density
$p_D(X,\phi)$ can be expressed as
\begin{equation}\label{eq20}
p_D(X,\phi)=-X+de^{\lambda\phi}X^{2}
\end{equation} being $d$ a positive constant and $X = \frac{1}{2}
\dot{\phi}^2$ the kinetic term of the dilaton scalar field $\phi$.
This is a higher-order kinetic correction to the usual kinetic
term motivated by dilatonic higher-order corrections to the
three-level action in low energy effective string theory
\cite{Piazza:2004df}. Since the $e^{\lambda \phi}$ term in Eq.~(\ref{eq20})
can be large for $\phi \to \infty$, the second term in
Eq.~(\ref{eq20}) can stabilise the vacuum
even if $X$ is much smaller than the Planck scale.

In string theory we have other non-perturbative and loop corrections such as the Gauss-Bonnet (GB) curvature invariant. Further, a dark energy model based on a string-inspired Lagrangian must in general contain higher derivative terms. It is also important to acknowledge the role that the GB coupling with the scalar field may play in the late-time universe \cite{Nojiri:2005vv,Nojiri:2006ri}. Moreover, the cosmological implications of the HDE density in the Gauss-Bonnet framework have been investigated in \cite{Setare:2008hm}.Therefore, it would certainly be of interest to extend my analysis to such a direction and this is left for a future work. However, in this paper, I have carried out the analysis for a simplified Lagrangian in order to understand the basic picture of the system. This seems to be justifiable \cite{Gasperini:2001pc}, and the dilatonic dark energy model obtained \cite{Piazza:2004df} possesses the characteristics of a viable model of dark energy.

We assume a spatially flat Friedmann-Robertson-Walker background
spacetime $ds^2 = dt^2 - a^2(t)\, d\vec{x}^2$ (where $a(t)$ is the
scale factor). Unless stated otherwise, we consider $\phi$ to be
smooth on scales of interest so that $X = \frac{1}{2}
\dot{\phi}^2\geq0$. The energy-momentum tensor of the dilaton is
obtained from Eq. (\ref{actiondilaton}), yielding
\begin{eqnarray}\label{energy-momentdilaton}
T_{\mu\nu}^{(\phi)}=-\frac{2}{\sqrt{-g}}\frac{\delta
S_{\phi}}{\delta
g^{\mu\nu}}=g_{\mu\nu}p_{D}+p_{,X}\partial_{\mu}\phi\partial_{\nu}\phi,
\end{eqnarray}
where $p_{,X} \equiv \partial p/\partial X$. Since the energy-momentum tensor
(\ref{energy-momentdilaton}) of the dilaton scalar field is that
of a perfect fluid, $T_{\mu \nu} = (\rho + p)u_\mu u_\nu + g_{\mu
\nu} p$, with velocity $u_\mu=\partial_\mu\phi/\sqrt{2X}$,
 we have the dilaton energy density $\rho_{D}$

\begin{eqnarray}\label{rhodilaton}
\rho_{D}=2Xp_{D,X}-p_{D}=-X+3de^{\lambda\phi}X^{2}
\end{eqnarray} and the Lagrangian density pressure in Eq.(\ref{eq20})
corresponds to the dilaton pressure $p_{D}$. Throughout this
paper, we will assume that the energy density is positive so that
$-X+3de^{\lambda\phi}X^{2}> 0$.

We know proceed to derive the stability conditions of the dilatonic dark energy by considering small fluctuations
$\delta\phi(t,\textbf{x})$ around a background value $\phi_{0}(t)$ which is the solution in the FRW spacetime. Then the field $\phi(t,\textbf{x})$ can be decomposed in the conventional form

\begin{eqnarray} \label{decompose}
\phi(t,\textbf{x})=
\phi_{0}(t)+\delta\phi(t,\textbf{x}) \,.
\label{phide}
\end{eqnarray} Since we are interested in ultra-violet (UV) instabilities, it is not restrictive to consider a Minkowski background.
Expanding $p_{D}(X,\phi)$ at the second order in $\delta\phi$, it is straightforward to find the Lagrangian and then the Hamiltonian for the fluctuations. The perturbed Hamiltonian reads

\begin{eqnarray}
{\cal H} &=& \left(p_{D,X}
+2 Xp_{D,XX}\right)
 \frac{(\delta \dot{\phi})^2}{2} \nonumber \\
& &
+p_{D,X}
\frac{(\nabla \delta \phi)^2}{2}
-p_{D,\phi \phi}
\frac{(\delta \phi)^2}{2}\,.
\label{Ldelpsi}
\end{eqnarray}

The positive definiteness of the Hamiltonian is guaranteed if the following conditions hold

\begin{eqnarray}
\label{xi}
\hspace*{-1.0em}
& &\xi_1\,  \equiv \,
p_{D,X} +2 X p_{D,XX}
\, \ge \, 0,~~
\xi_2 \, \equiv \,
p_{D,X}
\, \ge \, 0, \\
\hspace*{-1.0em}
& &\xi_3 \, \equiv \,
-p_{D,\phi \phi}
\, \ge \, 0\,.
\label{xi2}
\end{eqnarray}

When discussing the stability of classical perturbations, the quantity often used is the speed of sound $c_{s}$ defined by \cite{Garriga:1999vw}

\begin{eqnarray}
 \label{sound}
 c_s^2 \equiv \frac{p_{D,X}}
 {\rho_{D,X}}
 =\frac{\xi_2}{\xi_1}\,.
\end{eqnarray} In cosmological perturbation theory $c_{s}^{2}$ appears as a coefficient of the $k^{2}/a^{2}$ term, where $k$ is the comoving wavenumber. While the classical fluctuations may be regarded as stable when $c_{s}^{2}>0$, the stability of quantum fluctuations requires both the conditions $\xi_1>0$ and $\xi_2 \ge 0$. These two conditions prevent an instability related to the presence of negative energy ghost states. If these conditions are violated, the vacuum is unstable under a catastrophic production of ghosts and photon pairs \cite{Carroll:2003st,Cline:2003gs}. The production rate from the vacuum is proportional to the phase space integral on all possible final states. Since only a UV cut-off can prevent the creation of modes of arbitrarily high energies, this is essentially a UV instability. In our model the $e^{\lambda\phi}$ appearing in the second term of the RHS
in Eq.(\ref{eq20}) can be large for $\phi\rightarrow\infty$, so
that such a term in Eq.(\ref{eq20}) can stabilise the vacuum even
if $X$ was much smaller than the Planck scale. In particular, since in our model $\xi_{1}=-1+6de^{\lambda\phi}X$ and $\xi_{2}=-1+2de^{\lambda\phi}X$, the quantum stability is ensured for $de^{\lambda\phi}X\geq1/2$. The equation of state for the
dilaton can be written as $p_{D} = w_{D}\rho_{D}$ which rearranged
gives the equation of state parameter
\begin{eqnarray}\label{wdilaton}
w_{D} = \frac{p_{D}}{\rho_{D}}=
\frac{dXe^{\lambda\phi}-1}{3dXe^{\lambda\phi}-1}.
\end{eqnarray} Hence we have $w_{D}\geq-1$ under the condition $de^{\lambda\phi}X\geq1/2$, which means that the phantom equation of state ($w_{D}<-1$) if not realised if we want the model to be quantum mechanically stable.

Let us study now the cosmological dynamics of the dilatonic dark energy model in the flat FRW background. As a matter fluid, with energy density $\rho_{m}$, we take both baryons and cold dark matter. The Einstein equations in this case are

\begin{eqnarray}
& & 3H^2=\rho_{D}+\rho_{m}\,, \\
& & 2\dot{H}=-(2Xp_{D,X}+\rho_{m})\,, \\
& & \dot{\rho}_{D}+3H(\rho_{D}+p_{D})=0\,,
\end{eqnarray} where from here onwards we set $M_{P}=1$. Inserting Eqs. (\ref{eq20}) and (\ref{rhodilaton}) in the above equations yields

\begin{eqnarray}
& & 3H^2=-\frac12 \dot{\phi}^2+\frac34 de^{\lambda \phi}
\dot{\phi}^4+\rho_{m}\,, \\
& & 2\dot{H}=\dot{\phi}^2-de^{\lambda \phi} \dot{\phi}^4
-\rho_{m}\,, \\
& & \ddot{\phi} (3de^{\lambda \phi} \dot{\phi}^2-1)
+3H\dot{\phi} (de^{\lambda \phi} \dot{\phi}^2-1)
+\frac34 d\lambda e^{\lambda \phi} \dot{\phi}^4=0\,. \nonumber \\
\end {eqnarray}

In order to study cosmological dynamics in the presence of the dilaton scalar field and a background fluid, it is convenient to introduce the following dimensionless variables

\begin{equation}
\label{quan}
x_{1} \equiv \frac{\dot{\phi}}{\sqrt{6}H}\,,~~~
x_{2} \equiv \frac{e^{-\lambda \phi/2}}{\sqrt{3}H}\,.
\end{equation} which can be written in an autonomous form

\begin{eqnarray}
\label{dxgho}
\frac{{\rm d}x_{1}}{{\rm d}N} &=&
\frac32 x_{1} \left[1+x_{1}^2(dY-1)
 \right] \nonumber \\
& & +\frac{1}{1-6dY} \left[3(2dY-1)x_{1}+
\frac{3\sqrt{6}}{2}\lambda d x_{1}^2 Y
\right]\,,\\
\label{dygho}
\frac{{\rm d}x_{2}}{{\rm d}N} &=&
-\frac{\sqrt{6}}{2}\lambda x_{1}x_{2}+\frac32 x_{2}
[1+
x_{1}^2
(dY-1)]\,,
\end{eqnarray} where $N={\rm ln}\,a$ is the number of $e$-foldings
which is convenient to use for the dynamics of dark energy and

\begin{equation}
Y \equiv \frac{x_{1}^2}{x_{2}^2}=Xe^{\lambda \phi}\,.
\label{Ytach}
\end{equation}

The equation of state and the fraction of the energy
density for the dilaton field can now be written as
\begin{eqnarray}
\label{wdi}
& & w_{D}=\frac{1-dY}{1-3dY}\,, \\
\label{Omedi}
& & \Omega_{D}=\frac{\rho_{D}}{3H^{2}}=
-x_{1}^2+3d\frac{x_{1}^4}{x_{2}^2}\,.
\end{eqnarray}

The condition for the stability of quantum fluctuations corresponds to $dY \ge 1/2$. The following fixed points are relevant for viable cosmological evolution:

(a) Matter point: $(x_{1},x_{2})=(0,1/2)$.
This satisfies $w_{D}=-1$, $\Omega_{D}=0$ and $\Omega_{m}=1$.

(b) Accelerated point: $(x_{1},x_{2})=(-\sqrt{6}\lambda f_{-}(\lambda)/4,1/2\, + \lambda^{2}f_{+}(\lambda)/16)$, where

\begin{equation}
f_{\pm}\equiv1\pm\sqrt{1+16/(3\lambda^{2})}\,.
\label{fplusminus}
\end{equation} This satisfies $w_{D}=(-8+\lambda^{2}f_{+}(\lambda))/(8+3\lambda^{2}f_{+}(\lambda))$, $\Omega_{D}=1$ and $\Omega_{m}=0$. The cosmic acceleration occurs for $-1 \leq w_{D}<-1/3$,
i.e., $1/2 \leq dY<2/3$. This corresponds to the condition
$0 \leq \lambda^2 f_+(\lambda)<8/3$, i.e.,

\begin{equation}
0\leq\lambda <\sqrt{6}/3.
\label{lambdaacc}
\end{equation} It can be shown that this accelerated point is stable for $0 \leq \lambda <\sqrt{3}$ \cite{Piazza:2004df}. Hence the stability of the accelerated point is ensured under the condition (\ref{lambdaacc}).

We also have other fixed points. For example, there is another accelerated point $(x_{1},x_{2})=(-\sqrt{6}\lambda f_{+}(\lambda)/4,1/2\,+ \lambda^{2}f_{-}(\lambda)/16)$, but this corresponds to the quantum instability region $dY<1/2$ (i.e. the phantom equation of state $w_{D}<-1$). During the matter era we also have the scaling solution with $(x_{1},x_{2})=(\sqrt{6}/(2\lambda),1)$, $\Omega_{D}=3/\lambda^{2}$, and $w_{D}=0$. However, the existence of a viable scaling matter era requires the condition $\lambda>\sqrt{3}$, which is not compatible with the condition (\ref{lambdaacc}).

We shall study the stability of the fixed points in the case of $d=1$. The eigenvalues of the matrix ${\cal M}$ were numerically evaluated in Ref. \cite{Gumjudpai:2005ry} and it was shown that the determinant of
the matrix ${\cal M}$ for the point $(x_{1},x_{2})=(-\sqrt{6}\lambda f_{+}(\lambda)/4,1/2\,+ \lambda^{2}f_{-}(\lambda)/16)$
is negative with negative real parts of
$\mu_1$ and $\mu_2$.
Hence this phantom fixed point is a stable spiral.
As already mentioned, the point (b) is a stable node for $0<\lambda<\sqrt{3}$, whereas
it is a saddle point for $\lambda>\sqrt{3}$.
This critical value $\lambda_{*}=\sqrt{3}$
is computed by setting the determinant of ${\cal M}$ to be zero.
The point $(x_{1},x_{2})=(\sqrt{6}/(2\lambda),1)$ is physically meaningful for $\lambda > \sqrt{3}$
because of the condition $\Omega_{\phi} < 1$,
and it is a stable node \cite{Gumjudpai:2005ry}.
Hence the point $(x_{1},x_{2})=(\sqrt{6}/(2\lambda),1)$ is stable when the point (b) is unstable
and vice versa. It was shown in Ref.~\cite{Tsujikawa:2006mw}
 that this property
holds for all scalar-field models which possess scaling solutions.
We recall that the point $(x_{1},x_{2})=(-\sqrt{6}\lambda f_{+}(\lambda)/4,1/2\,+ \lambda^{2}f_{-}(\lambda)/16)$ is not stable at the quantum level.
The above discussion shows that the only viable attractor
which satisfies the conditions
of an accelerated expansion and the quantum stability
is the point (b).
Finally, we recall the sound speed of the dilatonic model is smaller that the speed of light because the condition $p_{,XX} \geq 0$ holds. The sound speed squared in this case is given by

\begin{equation}
c_{s}^{2}=\frac{2dY-1}{6dY-1}.
\end{equation} The condition (\ref{lambdaacc}) for the existence of the late-time accelerated point gives $1/2 \leq dY<2/3$. Hence the sound speed runs in the interval
\begin{equation}
0 \leq c_{s}<1/3
\end{equation} which means that this model does not violate causality.

We shall proceed with our study in
the light of the HDE with $c\geq 1$ as the
future event horizon is only well defined when $w_{D} \geq -1$ (see
\cite{Li:2004rb}) and we also want to ensure quantum stability.

In order to build our holographic model, we impose the holographic
nature to the dilatonic dark energy, i.e., we identify $\rho_{D}$
with $\rho_{\Lambda}$.

We consider a universe filled with a matter component $\rho_{m}$ and a holographic
dilatonic component $\rho_{D}$, the Friedmann equation (18)
can be equivalently expressed as
\begin{eqnarray}  \label{Friedmann1}
H(z)= H_{0}\left ( \frac{\Omega_{m,0}(1+z)^{3}}{1-\Omega_{D}}
\right )^{1/2}
\end{eqnarray}
where $z=(1/a)-1$ is the redshift of the universe. From the
definition of the HDE and the definition of
the future event horizon, we find
\begin{eqnarray}  \label{holo}
\int_{a}^{\infty}\frac{da^{\prime}}{Ha^{\prime
2}}=\int_{x}^{\infty}\frac{dx}{Ha}= \frac{c}{\sqrt{\Omega_{D}}Ha}
\end{eqnarray}
The Friedmann equation (\ref{Friedmann1}) implies
\begin{eqnarray}  \label{Ha}
\frac{1}{Ha}=\sqrt{a(1-\Omega_{D})} \frac{1}{H_{0}\sqrt{
\Omega_{m,0}}}
\end{eqnarray}
Inserting~(\ref{Ha}) into~(\ref{holo}), we arrive at
\begin{eqnarray}  \label{Ha1}
\int_{x}^{\infty} e^{x'/2} \sqrt{1-\Omega_{D}}dx'=ce^{x/2}
\sqrt{\frac{1}{\Omega_{D}}-1},
\end{eqnarray}
where $x=\ln a$. The differential equation for the fractional
density of dark energy is obtained by taking the derivative with
respect to $x$ in both sides of equation~(\ref{Ha1}), yielding
\begin{eqnarray}  \label{Omegadez}
\Omega_{D}^{'}=-(1+z)^{-1}\Omega_{D}(1-\Omega_{D}) \left (1+
\frac{2}{c}\sqrt{\Omega_{D}}\right ),
\end{eqnarray}
where the prime denotes the derivative with respect to the
redshift $z$. This equation has an exact solution [15] and
describes the evolution of the HDE as a
function of the redshift. Since $\Omega_{D}^{'}$ is always
positive, the fraction of dark energy increases with time. From
the energy conservation equation of dark energy, the equation of
state parameter of dark energy can be expressed as [15]
\begin{eqnarray}  \label{eos}
\omega_{D}=-1-\frac{1}{3} \frac{d \ln \rho_{D}}{d \ln a}
=-\frac{1}{3} \left (1+ \frac{2}{c}\sqrt{\Omega_{D}}\right ).
\end{eqnarray}
Note that the formula $\rho_{D}={\Omega_{D}\over
1-\Omega_{D}}\rho_{m,0}a^{-3}$ and the differential equation
of $\Omega_{D}$, Eq.(\ref{Omegadez}), are used in the second equal
sign.

The use of Eqs. (\ref{Friedmann1}),(\ref{rhodilaton}) and
(\ref{wdilaton}) allows the derivation of the kinetic term $X$ in
terms of holographic quantities
\begin{eqnarray}\label{Xofz}
\frac{X}{\rho_{cr,0}}=\frac{\Omega_{D}\Omega_{m,0}(1-3w_{D})(1+z)^{3}}{2(1-\Omega_{D})},
\end{eqnarray}
where $\Omega_{D}$ and $w_{D}$ are given by Eqs.(\ref{Omegadez})
and (\ref{eos}) respectively, and
$\rho_{cr,0}=3H_{0}^{2}$ is the critical density at the
present epoch.

Moreover, from the definition of the kinetic term $X = \frac{1}{2}
\dot{\phi}^2$ and Eq. (\ref{Xofz}), we can deduce derivative of
the holographic dilatonic scalar field $\phi$ with respect to the
redshift $z$

\begin{eqnarray}\label{phiofz}
\phi'=\mp \frac{\sqrt{3\Omega_{D}(1-3w_{D})}}{1+z},
\end{eqnarray} where the sign is in fact arbitrary as it can be changed by a
redefinition of the field $\phi\rightarrow-\phi$. The evolutionary
form of the holographic dilatonic field can be easily obtained
integrating the above equation numerically from $z=0$ to a given
value $z$. The field amplitude at the present epoch ($z=0$) is
taken to vanish, $\phi(0)=0$. Changing this initial value is
equivalent to a displacement in $\phi$ by a constant value
$\phi_{0}=\phi(z=0)$, which does not affect the shape of the
field.

   %%%%%%%%%%%%%%%%%%%%%%%%%%%%%%%%%%%%%%%%%%%%%
\begin{figure}
\psfrag{X}{$X$} \psfrag{c}{$c$} \psfrag{z}{$z$}
\includegraphics[scale=0.76]{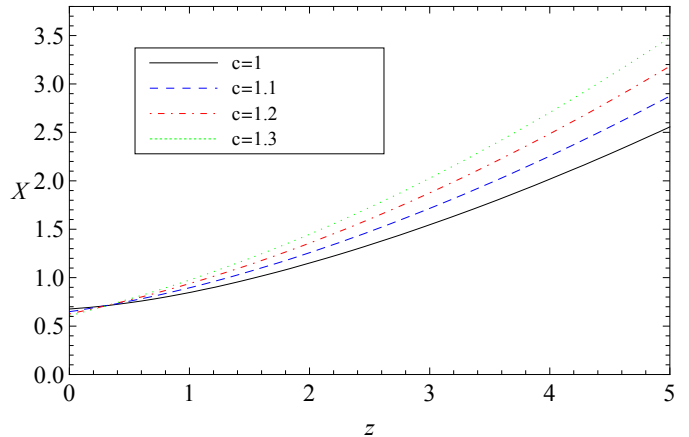}
\caption{Variation of $X(z)$, where $X$ is in units of
$3H_{0}^{2}$. We take here $\Omega_{m,0}=0.27$ and show
the cases for $c=1,1.1,1.2,1.3.$}

\end{figure}
%%%%%%%%%%%%%%%%%%%%%%%%%%%%%%%%%%%%%%%%%%%%%%%%

The parameter $c$ plays an essential role in describing the
evolution of the HDE model and should be determined by
cosmological observations. From Eq. (\ref{eos}) we see that the
equation of state parameter satisfies $-(1+2/c)/3 \leq w_{D} \leq
-1/3$ due to $0 \leq \Omega_{D} \leq 1$. If $c = 1$, the dark
energy equation of state parameter would asymptote to that of a
cosmological constant and the Universe would enter the de Sitter
phase in the future; if $c
> 1$, the equation of state parameter of dark energy would always
be greater than $-1$, behaving as quintessence dark energy; if $c
< 1$, the equation of state parameter of HDE would be initially
greater than $-1$, but it would decrease and eventually cross the
"phantom divide line" ($w_{D} = -1$) as the Universe expands, acting
as a quintom.

%%%%%%%%%%%%%%%%%%%%%%%%%%%%%%%%%%%%%%%%%%%%%
\begin{figure}
\psfrag{phi}{$\phi$} \psfrag{c}{$c$} \psfrag{z}{$z$}
\includegraphics[scale=0.70]{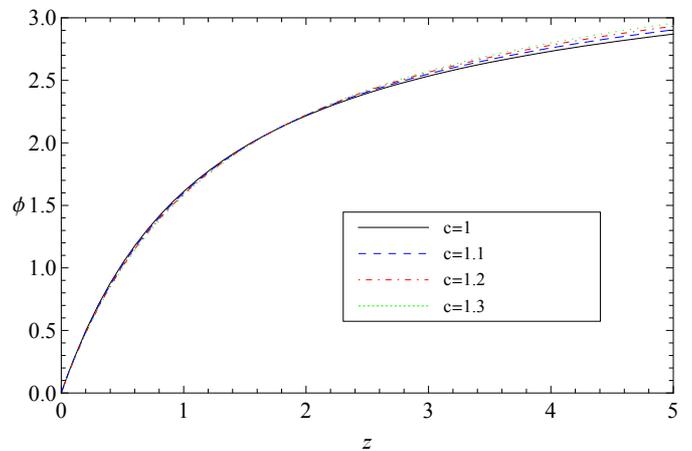}
\caption{The evolution of the dilaton scalar field $\phi(z)$ with
the $(-)$ sign in Eq.(\ref{phiofz}). We take here $\Omega_{m,0}=0.27$ and show the cases for
$c=1,1.1,1.2,1.3.$}

\end{figure}
%%%%%%%%%%%%%%%%%%%%%%%%%%%%%%%%%%%%%%%%%%%%%%%%

%%%%%%%%%%%%%%%%%%%%%%%%%%%%%%%%%%%%%%%%%%%%%
\begin{figure}
\psfrag{phi}{$\phi$} \psfrag{c}{$c$} \psfrag{z}{$z$}
\psfrag{X}{$X$}
\includegraphics[scale=0.615]{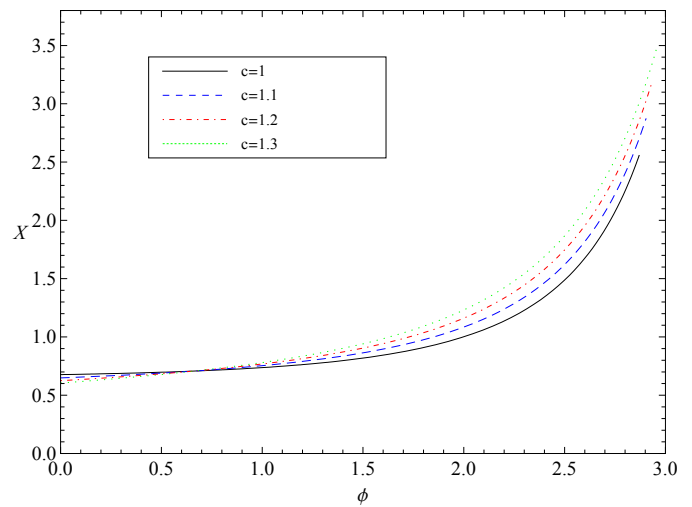}
\caption{Reconstructed $X$ for the holographic dilaton where $X$ is in units of $3H_{0}^{2}$. We
take here $\Omega_{m,0}=0.27$ and show the cases for
$c=1,1.1,1.2,1.3.$}

\end{figure}
%%%%%%%%%%%%%%%%%%%%%%%%%%%%%%%%%%%%%%%%%%%%%%%%

The best-fit analysis on the HDE model, by using the latest
observational data including the Union+CFA3 sample of 397 Type Ia
supernovae (SNIa), the shift parameter of the cosmic microwave
background (CMB) given by the five-year Wilkinson Microwave
Anisotropy Probe (WMAP5) observations, and the baryon acoustic
oscillations (BAO) measurement from the Sloan Digital Sky Survey
(SDSS) favors "quintom" behavior slightly. However,
quintessence-like behavior is also still allowed with the present
data \cite{Ma:2007pd}, \cite{Li:2009bn}. That is why the case
$c\geq1$ is worth investigating in detail. In addition,
\cite{Ma:2007pd} shows that $c < 1.2$ at more than $3\sigma$ which
is consistent with the possible theoretical limit of the parameter
$c$ from the weak gravity conjecture (see \cite{Ma:2007av}). It
was also found that the HDE model fits mildly better than the
$\Lambda$CDM, but with the data available at present the
difference is not significant.

The holographic evolution of the kinetic term can be obtained
numerically and it is shown in Fig.1 where we can see that $X$ is
a positive and monotonically increasing function with $z$ for an
accelerating Universe with HDE. Likewise, the
behavior of $\phi(z)$, obtained through Eq.(\ref{phiofz}), is
displayed in Fig.2.

The holographic dilatonic dark energy, represented by $X$, is
plotted in Fig.3 as a function of $\phi$. From Figs. 2 and 3 we
can see the dynamics of the field explicitly. Selected curves are
plotted for the cases of $c = 1.0, 1.1, 1.2$ and $1.3$, and the
present fractional matter density is chosen to be $\Omega_{m,0} =
0.27$. Given that the kinetic term decreases gradually with the
cosmic evolution, the equation of state parameter of the dilaton
$w_{D}$ tends to negative values close to $-1$ according to Eq.
(\ref{eos}) as $\dot{\phi}\rightarrow0$. As a result
$dw_{D}/d$lna$ <0$. Note that $\phi(z)$ increases with $z$ but
becomes finite at high redshift. This means that $\phi$ decreases
as the universe expands. Similar behavior was obtained in
\cite{holoquint} for the holographic quintessence and in
\cite{holotachyon} for the holographic tachyon model. Another
paper \cite{holoquintom} also dealt with the holographic dilaton
but the starting point, objectives, contents and conclusions in it
are different from ours.
\section{Conclusions}
\label{sec:2}

We have proposed a holographic dilatonic model of
dark energy with the future event horizon as infrared cut-off.
This has been done by establishing a correspondence between the
HDE model and the dilaton field. We have also carried out a
detailed analysis of its evolution and explore its cosmological
consequences.

By assuming that the scalar field models of dark energy are
effective theories of an underlying theory of dark energy and
regarding the scalar field model as an effective description of
such a theory, we can use the dilaton scalar field model to mimic
the evolving behavior of the HDE. As a result, we have
reconstructed the holographic dilatonic model in the region
$-1<w_{D}<-1/3$, which is the allowed region for this model when
$c\geq1$ and that is also quantum mechanically stable.

Therefore, we have shown that the holographic evolution of the
universe can be described completely by a dilaton scalar field for
$c\geq1$.

\end{document}